\documentclass[12pt]{iopart}
\usepackage{iopams}  
\usepackage{amssymb}
\usepackage{graphicx}
\usepackage{color}
\usepackage{bm}
\usepackage[normalem]{ulem}
\usepackage{dcolumn}
\usepackage{braket}
\usepackage[perpage]{footmisc}
\begin{document}

\title{Boundary Conditions and Vacuum Fluctuations in $\mathrm{AdS}_4$.}

\author{Vitor S. Barroso}
\address{Instituto de F\'isica ``Gleb Wataghin''\\
Universidade Estadual de Campinas - UNICAMP\\
13083-859, Campinas, SP, Brasil
}
\ead{barrosov@ifi.unicamp.br}

\author{J. P. M. Pitelli}%
\address{Instituto de Matem\'atica, Estat\'istica e Computa\c{c}\~ao Cient\'ifica\\
Universidade Estadual de Campinas - UNICAMP\\
13083-859, Campinas, SP, Brasil}%
\address{Also at The Enrico Fermi Institute, The University of Chicago, Chicago, IL}
 \ead{pitelli@ime.unicamp.br}
%
%
%
\vspace{10pt}
\begin{indented}
\item[]April 2019
\end{indented}

\begin{abstract}
Initial conditions given on a spacelike, static slice of a non-globally hyperbolic spacetime may not define the fates of classical and quantum fields uniquely. Such lack of global hyperbolicity is a well-known property of the anti-de Sitter solution and led many authors to question how is it possible to develop a quantum field theory on this spacetime. Wald and Ishibashi took a step towards the healing of that causal issue when considering the propagation of scalar fields on AdS. They proposed a systematic procedure to obtain a physically consistent dynamical evolution. Their prescription relies on determining the self-adjoint extensions of the spatial component of the differential wave operator. Such a requirement leads to the imposition of a specific set of boundary conditions at infinity. We employ their scheme in the particular case of the four-dimensional AdS spacetime and compute the expectation values of the field squared and the energy-momentum tensor, which will then bear the effects of those boundary conditions. We are not aware of any laws of nature constraining us to prescribe the same boundary conditions to all modes of the wave equation. Thus, we formulate a physical setup in which one of those modes satisfy a Robin boundary condition, while all others satisfy the Dirichlet condition. Due to our unusual settings, the resulting contributions to the fluctuations of the expectation values will not respect AdS invariance. As a consequence, a back-reaction procedure would yield a non-maximally symmetric spacetime. Furthermore, we verify the violation of weak energy condition as a direct consequence of our prescription for dynamics. 

\end{abstract}

%
%
%
\maketitle
%
%

\section*{\label{sec:intro}Introduction}

One of the most remarkable outcomes of string theory was the proposition of the $\mathrm{AdS/CFT}$ correspondence~\cite{maldacena}. It is conjectured that a theory of quantum gravity on $n$-dimensional $\mathrm{AdS}$ displays an underlying equivalent conformal quantum field theory without gravity, taking place at the $(n-1)$-dimensional conformal boundary of $\mathrm{AdS}$. Accordingly, applications to high energy and condensed matter physics appeared within the efforts to test the limits of this new conjecture, placing the anti-de Sitter spacetime under the scientific spotlight. 

Although most of the developments in $\mathrm{AdS}$ rely on string theory techniques, on a recent work~\cite{kent}, the authors have focused on studying semiclassical properties of the spacetime. Using the mathematical apparatus of Quantum Field Theory (QFT) in curved spaces, they have found the fluctuations of the expectation values of the energy-momentum tensor and the field squared in $\mathrm{AdS}_n$. However, they did not discuss in depth the implications of the causal structure of the spacetime, i.e., the effects of non-globally hyperbolicity. 

Since $\mathrm{AdS}$ has a conformal boundary, we may not be able to determine much about the history of a physical quantity without specifying its behavior at infinity. Such a circumstance poses a fundamental issue on the quantization procedure, namely the solutions of the wave equation will not be uniquely defined by initial conditions in $\mathrm{AdS}$, i.e., the Cauchy problem is not well-posed. Thus, unless we give extra information at the conformal boundary, the lack of predictability makes it impracticable to build a quantized field whose dynamical evolution comprises the entire history of the spacetime. 

Avis, Isham, and Storey~\cite{avis} were the first ones to address the causal pathology of AdS when solving field equations. They developed QFT  on $\mathrm{AdS}_4$ by regulating information leaving or entering the spacetime \emph{by hand}. Their approach proposes the imposition of boundary conditions at the spatial infinity in order to control whether information flows through (or is reflected by) the conformal boundary. Even though Avis et al. provide us with physically consistent solutions to the wave equation, works by Wald~\cite{wald} and Ishibashi~\cite{ishibashi1,ishibashi2} reveal that a broader category of boundary conditions might be employed to obtain a physical dynamical evolution.

In~\cite{ishibashi1}, the authors present a prescription for dynamics of fields in general non-globally hyperbolic spacetimes based on the grounds of physical consistency. In order to fulfill some reasonable physical requirements (to be explained later),  they argue that the spatial component of the differential wave operator must be self-adjoint. Besides, in~\cite{ishibashi2}, they show that the prescription for dynamics in $\mathrm{AdS}$ translates into specifying boundary conditions at the conformal boundary. While Kent and Winstanley, in~\cite{kent}, impose the Dirichlet boundary condition at infinity, perhaps without realizing, they are neglecting an entire set of non-equivalent dynamical outcomes. According to Ishibashi and Wald~\cite{ishibashi2}, those outcomes would correspond to the various boundary conditions that one could have specified at infinity.

In this paper, we study physical effects that may arise due to non-Dirichlet boundary conditions at the conformal boundary.   We investigate those effects by computing the vacuum fluctuations of the expectation values of the quadratic field and the energy-momentum tensor for conformally coupled scalar fields in $\mathrm{AdS}_4$. Also, we will keep Ref.~\cite{kent} as a basis for our results and shall return to it for further comparison.  

We have organized this article as follows. In Sec.~\ref{sec:ads}, we briefly review some of the fundamental aspects of the anti-de Sitter solution. Then, in Sec.~\ref{sec:ngh}, we display the systematic procedure that describes the dynamics of scalar fields in non-globally hyperbolic spacetimes - such as $\mathrm{AdS}$ - first presented by Wald and Ishibashi. With that scheme in hands, we show the implications their prescription has on scalar fields propagating on $\mathrm{AdS}$, in Sec.~\ref{sec:scalarads}. Our next step is to build the proper Green's functions in Sec.~\ref{sec:greenads}, and employ them in the computations of the renormalized quantities of interest, namely the fluctuations of the expectation values of the field squared and the energy-momentum tensor, both shown in Sec.~\ref{sec:renormads}. Finally, we discuss our results in Sec.~\ref{sec:disc}.

\section{\label{sec:ads} Anti-de Sitter spacetime}

Surfaces of constant negative curvature are well-known in geometry and comprise the set of hyperbolic spaces. In the context of General Relativity, the equivalent to those spaces is the $n$-dimensional anti-de Sitter space, which appears as a solution to Einstein equations when choosing a negative cosmological constant ($\Lambda <0$) in the absence of matter and energy. Setting $\Lambda := -\frac{(n-1)(n-2)}{2 H^2} $, we may write the Einstein equations as
\begin{equation}
R_{\mu\nu}-\frac{1}{2}R g_{\mu\nu}-\frac{(n-1)(n-2)}{2 H^2} g_{\mu \nu}=0.
\end{equation}
The outcome is an $n$-dimensional maximally symmetric pseudo-Riemmanian metric defined over a Lorentzian manifold with constant negative curvature, i.e., the $\mathrm{AdS}_n$ spacetime. In a suitable set of parametrized coordinates $\{x^\mu\}$\footnote{ The radial coordinate, $\rho$, is defined over the interval $[0,\pi/2)$. The polar and azimuthal coordinates on the unit $(n-2)$-sphere are $\theta_j \ (j=1,\dots,n-3)$ and $\varphi:=\theta_{n-2}$, respectively, each satisfying  $0\leq \theta_j \leq\pi$ and $0\leq\varphi<2\pi$. The timelike coordinate, $\tau$, ranges from $-\pi$ to $\pi$.}, the line element for the induced metric $g_{\mu\nu}$ on $\mathrm{AdS}_n$ is
\begin{equation}
ds^2= H^2(\sec\rho)^2[-d\tau^2+d\rho^2+(\sin\rho)^2 d\Omega_{n-2}^2],
\label{eq:adsmetric}
\end{equation}
where $d\Omega_{n-2}^2$ is the line element on a unit $(n-2)$-sphere.

\subsection{\label{sec:topologyads}Topology}
We may understand $\mathrm{AdS}_n$ as an isometric embedding of a single sheeted $n$-dimensional hyperboloid in an $(n+1)$-dimensional flat space provided with metric $\mathrm{diag}(-1,1,\cdots1,-1)$. Timelike curves in $\mathrm{AdS}$ are transverse sections of the hyperboloid, and they are always closed. The periodicity of the timelike coordinate, $\tau$, suggests that given a point in spacetime, we can return to it by only traveling along a timelike geodesic of length $2\pi$ in $\tau$. Accordingly, the topology of $\mathrm{AdS}_n$ becomes apparent, namely $\mathbb{S}^1\times\mathbb{R}^{n-1}$, which is compatible with the existence of closed timelike curves. Thus, unphysical events can take place in the spacetime, such as a particle returning to the same position through a periodic motion in time.

\subsection{\label{sec:causalads}Causal structure}

Wald remarks in~\cite{wald1} that observers following closed timelike geodesics would have no difficulty altering past events hence breaking causality. In an attempt to solve this primary issue, we can 'unwrap' the hyperboloid along the timelike direction, and patch together unwrapped hyperboloids one after the other. In other words, we construct a spacetime spatially identical to $\mathrm{AdS}$ but extended in time, i.e., the temporal coordinate no longer ranges from $-\pi$ to $\pi$ but from $-\infty$ to $\infty$. We refer to such procedure as the universal covering of $\mathrm{AdS}$, and the resulting spacetime as $\mathrm{CAdS}$. 

Even though the unwrapping of $\mathrm{AdS}$ prevents the existence of closed timelike curves, another fundamental causality issue remains, namely the lack of predictability associated with fields propagating on the spacetime. Indeed, no Cauchy hypersurfaces exist in $\mathrm{AdS}$ (and $\mathrm{CAdS}$) hence portraying it as a non-globally hyperbolic spacetime. The Cauchy problem will not be well-posed, yielding non-unique dynamics for a given set of initial conditions. We can understand this scenario as a result of information leaking through the spatial infinity of the spacetime, i.e., flowing in (out) from (through) the boundary. In order to solve such a pathological behavior, we shall discuss in the next sections how to adequately address causality issues associated with field equations in non-globally hyperbolic spacetimes.

\section{\label{sec:ngh}Scalar fields in non-globally hyperbolic static spacetimes}

An extensive literature (see, for instance, \cite{birrel} and references therein) provides a complete guide on  QFT in curved spaces, and conduct us through a generalized quantization procedure based on that of QFT in Minkowski spacetime. Nevertheless, several researchers developed most of it in a category of spacetimes whose causal structure is thoroughly well-defined, namely globally hyperbolic spacetimes. Indeed, as we discussed previously if a spacetime does not feature global hyperbolicity, then basic field equations might not have causal solutions, which jeopardizes the quantization of fields. On what follows, we use works by Wald~\cite{wald} and Ishibashi~\cite{ishibashi1,ishibashi2} to prescribe the appropriate dynamics of scalar fields in non-globally hyperbolic spacetimes.

Let us consider a static spacetime $(\mathcal{M},g_{\mu\nu})$, which admits the following decomposition of its metric~\cite{hawking}
\begin{equation}
    ds^2=g_{\mu \nu}dx^\mu dx^\nu=-V^2dt^2+h_{ij}dx^idx^j.
    \label{eq:static}
\end{equation}
In Eq.~\ref{eq:static}, $h_{ij}$ is the metric induced on a hypersurface $\Sigma$ orthogonal to a given timelike Killing field $\tau^\mu$ of the metric, and we define $V^2=-\tau^\mu \tau_\mu$. In this particular case, Klein-Gordon equation, 
\begin{equation}
    \nabla_\mu \nabla^\mu \phi - m^2 \phi - \xi R \phi = 0, \label{eq:kg}
\end{equation}
reduces to
\begin{equation}
    \partial_t^2 \phi = - A \phi,   
    \label{eq:wave}
\end{equation}
in which $A := -V D^i(VD_i \phi)+m^2 V^2+\xi R V^2$ is the spatial component of the wave operator, and $D_i$ is the covariant derivative in a spatial slice of $\Sigma$.

Wald points out in~\cite{wald} that $A$ is an operator defined on a Hilbert space $\mathcal{H}=\mathcal{L}^2(\Sigma)$ with domain $\mathcal{D}(A)=C_0^\infty(\Sigma)$, and whose self-adjointness properties are relevant to examine the dynamical evolution appropriately. An extensive literature on Functional Analysis (e.g., see \cite{reed1,reed2}) discusses the properties of such operators and present a systematic procedure for obtaining their self-adjoint extensions, accredited to Weyl and von Neumann.

It can be easily checked that $(A,\mathcal{D}(A))$ defined above is symmetric. For such a symmetric  operator, we denote by $(A^\dagger,\mathcal{D}(A^\dagger))$ its adjoint operator. Symmetry of $A$ implies that $A=A^\dagger$. However, we may have $\mathcal{D}(A)\neq \mathcal{D}(A^\dagger)$ - when $A$ is not self-adjoint. In this case it may be possible to find the self-adjoint extensions of $A$. In order to find these extensions, let us  define the \emph{deficiency subspaces} of $A$, denoted $\mathcal{N}_\pm\subset\mathcal{H}$, by
\begin{equation}
    \mathcal{N}_\pm=\{\psi_\pm\in\mathcal{D}(A^\dagger)|\  A^\dagger\psi_\pm=\pm i\lambda\psi_\pm, \lambda\in\mathbb{R}^+\},\label{eq:defsub}
\end{equation}
and the \emph{deficiency indices} as $n_\pm=\dim (\mathcal{N}_\pm)$. There are three cases to be considered:
  \begin{enumerate}
        \item If $n_+\neq n_-$, then $A$ has no self-adjoint extension.
        \item If $n_+=n_-=0$, then $A$ is essentially self-adjoint, and we obtain it by taking the closure, $\bar{A}$, of $A$.
        \item If $n_+=n_-=n\geq 1$, then infinitely many self-adjoint extensions of $A$ may exist. They are in one-to-one correspondence to the isometries between $\mathcal{N}_+$ and $\mathcal{N}_-$ parametrized by an $n\times n$ unitary matrix, $U$.
    \end{enumerate}
Certainly, the third case is more complex than the others, and we must follow a method for obtaining the self-adjoint extensions (see~\cite{reed2} for a proper description of it). They are given by $A_E$, with $E$ being a parameter labeling the extension, defined by
\begin{equation}
    \mathcal{D}(A_E)=\{\Phi_0+\Phi_+ +U\Phi_+ | \ \Phi_0\in\mathcal{D}(A),\Phi_+\in\mathcal{N}_+\},\label{eq:domain}
\end{equation}
and 
\begin{equation}
    A_E \Phi=A\Phi_0+i\Phi_+-i U \Phi_+,
\end{equation}
for all $\Phi\in\mathcal{D}(A_E)$. This procedure can always be followed to find whether an operator has self-adjoint extensions and identify them, in case they exist.

In particular, Wald~\cite{wald} proposes that there might exist a set of solutions of the wave equation \ref{eq:wave} associated with each self-adjoint extension, i.e.,
\begin{equation}
    \phi_t=\cos(A_E^{1/2}t)\phi_0+A_E^{-1/2}\sin(A_E^{1/2}t)\dot{\phi}_0, \label{eq:din}
\end{equation}
given well-posed initial conditions to the Cauchy problem, namely $(\phi_0,\dot{\phi}_0)\in C_0^\infty(\Sigma)\times C_0^\infty(\Sigma)$, for all $t\in \mathbb{R}$.

It is straightforward to notice that for each extension $A_E$ there will be an associated dynamical evolution of Eq.~\ref{eq:din}. Consequently, the dynamics of the field is not uniquely determined by initial conditions. We identify those non-equivalent solutions as a result of various boundary conditions that one can impose at a region in space, such as a singularity or a boundary~\cite{wald}. Ishibashi and Wald, in \cite{ishibashi1}, argue that Eq.~\ref{eq:din} is the only one that prescribes a physically sensible dynamics of scalar fields in non-globally hyperbolic static spacetimes. By comparison with the globally hyperbolic case, they establish a set of conditions that determine whether a time evolution is consistent or not, namely:
\begin{enumerate}
    \item solutions of the wave equation must be causal;
    \item the prescription for dynamics must be invariant under time translation and reflection;
    \item there exists a conserved energy functional also respecting time translation and reflection invariance, in agreement with the globally-hyperbolic case;
    \item solutions satisfy a convergence condition, as proposed in \cite{wald}.
\end{enumerate}

\section{\label{sec:scalarads} Boundary conditions at infinity of anti-de Sitter}

Let us now consider Klein-Gordon equation \ref{eq:wave} in $\mathrm{AdS}_n$, as follows
\begin{equation}
    \partial_t^2\phi=-(\sec\rho)^2\Big\{(\cot\rho)^2\Big[-(n-2)\tan\rho\partial_\rho^2-\Delta_S\Big]-H^2 m_\xi^2\Big\}\phi,\label{eq:adswave}
\end{equation}
where $m_\xi$ is the effective mass of the field defined by $m_\xi^2=m^2-\xi n(n-1)H^{-2}$, and
\begin{equation}
     \Delta_S=\sum_{j=1}^{n-3}\Big[(n-2)\cot\theta_j\partial_{\theta_j}+\prod_{k=1}^{j-1}(\csc\theta_k)^2\partial^2_{\theta_j}\Big]
+\prod_{j=1}^{n-3}(\csc\theta_j)^2\partial^2_{\varphi}
\end{equation}
is the Laplace-Beltrami operator on the unit $(n-2)$-sphere whose eigenfunctions are Generalized Spherical Harmonic functions, $Y_l(\theta_j,\phi)$, with eigenvalues $l(l+n-3)$. We may recall that a static slice of $\mathrm{AdS}_n$ can be decomposed into a real interval $[0,\pi/2)$, labeled by the radial coordinate $\rho$, and an $(n-2)$-dimensional unit sphere $\mathbb{S}^{n-2}$, parametrized by the angular coordinates $\theta_j$ and $\varphi$. It is also worth pointing out that, as the spacetime is static, there exists a timelike Killing field $\partial_t$, whose eigenfunctions $e^{-i\omega t}$ with positive energy, $\omega>0$, can be used to expand the solution $\phi$. Thus, $\phi$ will be an eigenfunction of the quadratic operator $\partial_t^2$ with eigenvalue $-\omega^2$. With those considerations in hand, let us write the solution as 
\begin{equation}
    \phi(t,\rho,\theta_j,\varphi) =\sum_{\omega, l}e^{-i\omega t}\tilde{f}_{\omega, l}(\rho)Y_l(\theta_j,\phi).
\end{equation}

Under the transformation 
\begin{equation}
    \tilde{f}_{\omega, l}(\rho)=(\cot\rho)^{\frac{n-2}{2}} f_{\omega, l}(\rho),\label{eq:conformradial}
\end{equation}
and omitting temporal and angular dependence, Eq.~\ref{eq:adswave} reduces to
\begin{equation}
    A f_{\omega, l}(\rho)=\omega^2 f_{\omega, l}(\rho), \label{eq:waveop}
\end{equation}
upon the identification~\cite{ishibashi2}\footnote{Ishibashi and Wald define the radial coordinate $x$ for the spatial infinity to be located at $x=0$\cite{ishibashi2}. It relates to our radial coordinate $\rho$ by $x=\pi/2-\rho$.}
\begin{equation}
    A\equiv-\frac{d^2}{d\rho^2}+\frac{\nu^2-1/4}{(\cos\rho)^2}+\frac{\sigma^2-1/4}{(\sin\rho)^2},\label{eq:aopads}
\end{equation}
which is a differential operator whose domain is $C_0^\infty(0,\pi/2)$ defined over a Hilbert space \mbox{$\mathcal{H}=\mathcal{L}^2([0,\pi/2],d\rho)$}, and the coefficients of the equation are defined as
\begin{equation}
    \nu^2-1/4=\frac{n(n-2)}{4}+H^2m^2-n(n-1)\xi,\label{eq:nu2}
\end{equation}
and
\begin{equation}
    \sigma^2-1/4=\frac{(n-2)(n-4)}{4}+l(l+n-3).\label{eq:sigma2}
\end{equation}
From Eq.~\ref{eq:sigma2}, it is straightforward to check that
\begin{equation}
    \sigma=l+\frac{n-3}{2}.
\end{equation}
The coefficient $\nu$ is taken to be the positive square root of $\nu^2$ and will depend on the mass and coupling factor of the field. In such conditions, there are four relevant cases to be analyzed, namely
\begin{itemize}
    \item[(i)] $\nu^2\geq 1$: in this case, the effective mass of the field satisfies the relation $H^2m_\xi^2\geq -(n+1)(n-3)/4$, which comprise the minimally coupled, massless scalar field for $n\geq 3$.
    
    \item[(ii)] $0<\nu^2<1$: this case occurs for $-(n-1)^2/4<H^2m^2_\xi< -(n+1)(n-3)/4$, and includes conformally invariant scalar fields in all dimensions.
    
    \item[(iii)] $\nu^2=0$: this is the case when the effective mass squared reaches a critical value, namely $H^2m^2_\xi\equiv -(n-1)^2/4$.
    
    \item[(iv)] $\nu^2<0$: in this case, the effective mass squared is lower than the critical mass, i.e., $H^2m^2_\xi< -(n-1)^2/4$.
\end{itemize}
In \cite{ishibashi2}, the authors examine the positivity of the operator $A$ in terms of $\nu$. They demonstrate that, in all cases in which $\nu^2\geq0$ - i.e., in $(i)$, $(ii)$ and $(iii)$ - $A$ is a positive operator. Meanwhile, in case $(iv)$, the operator is unbounded bellow. Consequently, $A$ has no positive, self-adjoint extensions in case $(iv)$. On the other hand, at least one self-adjoint extension to $A$ exists - that is, the Friedrichs extension\cite{reed1} - in all other cases: $(i)$, $(ii)$ and $(iii)$.

The solutions to Eq.~\ref{eq:waveop} are  given by
\begin{eqnarray}
	 f_{\omega,l}(\rho)&=\mathbf{C}\cdot(\cos \rho)^{\nu+1/2}\cdot(\sin \rho)^{\sigma+1/2}\nonumber\\ 
	 &\qquad \times _2F_1\left(\frac{\nu+\sigma+\omega+1}{2},\frac{\nu+\sigma-\omega+1}{2};1+\sigma,(\sin \rho)^2\right).\label{eq:solOpA}
\end{eqnarray}
The other linear independent solution is never square-integrable, so we neglect it here. According to Eq.~\ref{eq:defsub}, to construct the deficiency subspaces $\mathcal{N}_\pm$, we must take $\omega^2=\pm\lambda i$, so $\omega\in\mathbb{C}$. In such conditions, as shown in~\cite{ishibashi2}, solution \ref{eq:solOpA} fails to be square integrable in case $(i)$, i.e., $\nu\geq 1$. However, for $0\leq\nu<1$, which corresponds to cases $(ii)$ and $(iii)$, $f$ is square integrable for all $\omega\in\mathbb{C}$.

In case $(i)$, the deficiency subspaces are trivial, so $n_+=n_-=0$, and the operator admits a unique self-adjoint extension. In other words, the repulsive effective potential in $A$, i.e., $(\cos\rho)^{-2}$, prevents the fields from reaching spatial infinity. Hence, they vanish there, and no additional boundary conditions are required. Conversely, in cases $(ii)$ and $(iii)$, the deficiency subspaces $\mathcal{N}_\pm$ are each spanned by an eigenfunction $f_\pm$ of $A$ with eigenvalue $\omega^2=\pm 2i$. Thus, the deficiency indices in these cases are $n_+=n_-=1$, so infinitely many positive self-adjoint extensions of $A$ exist. Now, the effective potential is not as strong as in case $(i)$; hence we may associate the extensions to boundary conditions prescribed at infinity.

%

A one-parameter family of self-adjoint extensions, $A_\beta$, of $A$ exists for $0\leq\nu^2<1$ (cases $(ii)$ and $(iii)$). Equation \ref{eq:domain} provides us with the appropriate domain of $A_\beta$. Since the domain of $A$ consists of functions in $C_0^\infty$, all additional information needed to prescribe a physically consistent dynamical evolution must come from the asymptotic behavior of $f_+$ and $Uf_+$, for all isometries $U$. 

Let $U_\beta$ denote the isometries between $\mathcal{N}_+$ and 
$\mathcal{N}_-$, given by
\begin{equation}
    U_\beta f_+ = e^{i\beta}f_-, 
\end{equation}
for $\beta\in(-\pi,\pi]$. Let us consider the function
\begin{equation}
    f_\beta:=f_++U_\beta f_+\equiv f_++e^{i\beta} f_-,
\end{equation}
whose behavior near infinity ($\rho=\pi/2$) dictates the boundary conditions satisfied by all solutions $\phi_t$ of the form \ref{eq:din}. For $0<\nu<1$, the asymptotic behavior at $\rho=\pi/2$ is
\begin{equation}
        f_\beta\propto(\sin\rho)^{\sigma+1/2}\cdot(\cos\rho)^{-\nu+1/2}\\
        \times (a_\nu+b_\nu (\cos\rho)^{2\nu}+c_\nu(\cos\rho)^2+\dots),
\end{equation}
where the coefficients of the leading terms, $a_\nu$ and $b_\nu$, are functions of $\nu$, $\sigma$, the spacetime dimension $n$ and the parameter $\beta$. The leading powers in $\rho$ of $f_+$ are
\begin{equation}
    f_\beta\approx b_\nu \left(\frac{\pi}{2}-\rho\right)^{\nu+1/2}\left\{1+\frac{a_\nu}{b_\nu}\left(\frac{\pi}{2}-\rho\right)^{-2\nu}\right\},\label{eq:asympf+}
\end{equation}
from which we can see that the asymptotic boundary condition depends on the ratio $a_\nu/b_\nu$, which may take any real value. For $\nu=0$, we have
\begin{equation}
 \fl f_\beta\propto(\sin\rho)^{\sigma+1/2}\cdot(\cos\rho)^{1/2}
 \times(a_0\log(\cos^2\rho)+b_0+c_0(\cos\rho)^2\log(\cos^2\rho)+\dots),
\end{equation}
and an analogous procedure reveals that the asymptotic boundary condition depends on $a_0/b_0$ also in this case. However, the function $(\sin\rho)^{-\sigma-1/2}\cdot(\cos\rho)^{-1/2}\cdot f_\beta$ and its first derivative in $\rho$ both scale with $a_0$ when approaching infinity $\rho=\pi/2$. Setting $a_0=0$, we recover  Dirichlet  and Neumann boundary condition imposed simultaneously, which is precisely Friedrichs extension.

On what follows, we shall denote the ratio $a_\nu/b_\nu$ by $\alpha_\nu$, hence all self-adjoint extensions of the operator will be parametrized by $\alpha$ instead of $\beta$, although $\alpha\equiv\alpha(\beta)$. From Eq.~\ref{eq:asympf+}, we can check that\footnote{We exchanged all indices $\beta$ for $\alpha$.}
\begin{equation}
   \left. \frac{\frac{d}{d\rho}[(\sin\rho)^{-\sigma-1/2}\cdot(\cos\rho)^{\nu-1/2}\cdot f_\alpha]}{[(\sin\rho)^{-\sigma-1/2}\cdot(\cos\rho)^{3\nu-3/2}\cdot f_\alpha]}\right|_{\rho=\pi/2}=-2\nu\frac{1}{\alpha_\nu},\label{eq:GenRobinbc}
\end{equation}
which we identify as generalized Robin boundary conditions for $0<\nu<1$. One recovers generalized Dirichlet or Neumann boundary conditions by setting $\alpha_\nu$ equals to $0$ and $\pm\infty$, respectively. In the particular case $\nu=1/2$, Eq.~\ref{eq:asympf+} reduces to an even simpler form of the boundary conditions given by\footnote{In case $\nu=1/2$, we drop the index of $\alpha_{1/2}$ and replace it simply by $\alpha$.}
\begin{equation}
    \left[\frac{df_\alpha}{d\rho}\Big/f_\alpha\right]_{\rho=\pi/2}=-\frac{1}{\alpha},\label{eq:Robinbc}
\end{equation}
which is the usual Robin boundary condition, hence mixing Dirichlet ($\alpha=0$) and Neumann ($\alpha=\pm\infty$) conditions.

Even though the extensions $A_\alpha$ are now parametrized by a real parameter $\alpha_\nu$, not all of them are positive. Except for $\nu^2\geq1$, whose unique self-adjoint extension is already positive, the remaining cases satisfy the positivity conditions shown in \cite{ishibashi2}:

For $0<\nu^2<1$, we have
\begin{equation}
    \frac{b_\nu}{a_\nu}\equiv\frac{1}{\alpha_\nu}\geq-\left|\frac{\Gamma(-\nu)}{\Gamma(\nu)}\right|\frac{\Gamma(\frac{\sigma+\nu+1}{2})^2}{\Gamma(\frac{\sigma-\nu+1}{2})^2}.\label{eq:positi}
\end{equation}

For $\nu^2=0$, we have
\begin{equation}
    \frac{b_0}{a_0}\leq 2\gamma+2\psi\left(\frac{\sigma+1}{2}\right),\label{eq:positi0}
\end{equation}
where $\gamma$ is the Euler gamma and $\psi$ is the digamma function.

It is worth pointing out that equations \ref{eq:GenRobinbc} and \ref{eq:Robinbc} must be satisfied mode by mode, i.e., for each spherical label $l$ - and for each $\sigma$, indirectly (see Eq.~\ref{eq:sigma2}) -, the conditions are satisfied by $f_{\beta,\omega,l}$. Accordingly, there are infinitely many parameters $\alpha_{\nu,l}$ associated to each $f_{\beta,\omega,l}$, and they all satisfy different positivity conditions, given in equations \ref{eq:positi} and \ref{eq:positi0}.

\section{Green's functions in $\mathrm{AdS}$ }\label{sec:greenads}

In~\cite{allen}, Allen and Jacobson show that, in a maximally symmetric spacetime, two-point functions such as $G_F(x,x')=-i\langle \psi|T\{\phi(x)\phi(x')\}|\psi\rangle$, where $|\psi\rangle$ is a maximally symmetric state, may be written in terms of the geodetic interval $s(x,x')$\footnote{In $\mathrm{AdS}$, $s$ is constructed so that it goes to zero as $x'\rightarrow x$ and goes to infinity as we approach the boundary}, i.e.,
\begin{equation}
    G_F(x,x'):= G_F(s(x,x'))\equiv G_F^{(AJ)}(s).
\end{equation}
Their proposition simplifies the computations considerably since the wave equation becomes an ODE of the variable $s$. They also require that the Green's function falls off as fast as possible at spatial infinity, which in $\mathrm{AdS}$ translates into: $G_F\rightarrow 0$ as $s\rightarrow\infty$. In other words, they are choosing Dirichlet boundary condition for the field $\phi$.  Kent and Winstanley, in~\cite{kent}, exploit this simplicity to find the fluctuation of the field squared and the energy-momentum tensor in all spacetime dimensions of $\mathrm{AdS}$. They also verify that their results are compatible with the ones of Burgess and L\"{u}tken, whose approach in~\cite{burgess} was to perform a summation of modes of the wave solutions.

We are not aware of any law of nature that restricts the boundary conditions of all modes to Dirichlet ones. Indeed, Ishibashi and Wald showed in \cite{ishibashi1} that there is an entire category of boundary conditions that prescribe a physically consistent dynamical evolution. Additionally, there is no guarantee that all modes must satisfy the same boundary condition.

Let us then consider a setup in which one of the modes of the wave equation, $u_{\omega_\alpha, l_\alpha}$, is chosen so that its radial component $f_{\omega_\alpha,l_\alpha}(\rho)$ satisfies a generalized Robin boundary condition with parameter $\alpha$. Meanwhile, the components $f_{\omega,l}(\rho)$ of all other modes $u_{\omega, l}(x)$ ($l\neq l_\alpha$) satisfy Dirichlet boundary  conditions. 

The Green's function in this case is given by mode sum (from now on, we consider $\tau>\tau'$)

\begin{eqnarray}
\fl G_F(x,x')=-i(\cot\rho\cot\rho')^{\frac{n-2}{2}}\nonumber
    \\ \quad\times\bigg\{\sum_{\omega_\alpha} |\mathcal{N}_{\omega_\alpha, l_\alpha}|^2Y_{l_\alpha}(\theta_j,\varphi)Y^*_{l_\alpha}(\theta_j',\varphi')f_{\omega_\alpha, l_{\alpha}}(\rho)f_{\omega_\alpha, l_\alpha}(\rho')e^{-i\omega_\alpha(\tau-\tau')}\nonumber
    \\\qquad\qquad+\sum_{\stackrel{l\geq0}{l\neq l_\alpha}}\sum_\omega |\mathcal{N}_{\omega, l}|^2Y_l(\theta_j,\varphi)Y^*_l(\theta_j',\varphi')f_{\omega, l}(\rho)f_{\omega, l}(\rho')e^{-i\omega(\tau-\tau')}\bigg\},\label{eq:greenmod1}
\end{eqnarray}

where $\mathcal{N}_{\omega,l}$ are normalization constants. We may complete the last term in the summation for all Dirichlet modes by adding them to and subtracting them off Eq.~\ref{eq:greenmod1}, i.e.,

\begin{eqnarray}
  \fl  G_F(x,x')=-i(\cot\rho\cot\rho')^{\frac{n-2}{2}}\nonumber\\
   \times\bigg\{\sum_{\omega_\alpha} |\mathcal{N}_{\omega_\alpha, l_\alpha}|^2Y_{l_\alpha}(\theta_j,\varphi)Y^*_{l_\alpha}(\theta_j',\varphi')f_{\omega_\alpha, l_\alpha}(\rho)f_{\omega_\alpha, l_\alpha}(\rho')e^{-i\omega_\alpha(\tau-\tau')}\nonumber\\
    \qquad-\sum_{\omega} |\mathcal{N}_{\omega, l_\alpha}|^2Y_{l_\alpha}(\theta_j,\varphi)Y^*_{l_\alpha}(\theta_j',\varphi')f_{\omega, l_\alpha}(\rho)f_{\omega, l_\alpha}(\rho')e^{-i\omega(\tau-\tau')}\nonumber
    \\\qquad \qquad+\sum_{l,\omega} |\mathcal{N}_{\omega, l}|^2Y_l(\theta_j,\varphi)Y^*_l(\theta_j',\varphi')f_{\omega, l}(\rho)f_{\omega, l}(\rho')e^{-i\omega(\tau-\tau')}\bigg\}.\label{eq:greenmod2}
\end{eqnarray}

Let us denote the last term in equation \ref{eq:greenmod2} by $G_F^{(D)}$, and the first two terms by $G_F^{(\alpha)}$. The Green's function $G^{(D)}$ is obtained by the summation of Dirichlet modes purely. Thus, in the coincidence limit, it recovers the same results as $G^{(BL)}$, by Burgess and L\"{u}tken, and $G_F^{(AJ)}$, by Allen and Jacobson. On the other hand, $G_F^{(\alpha)}$ lacks contributions from all spherical components, since it is not summed over all angular modes $l$. Hence, $G_F^{(\alpha)}$ may not be a maximally symmetric function. It seems reasonable for us to write that
\begin{equation}
    G_F(x,x')\equiv G_F^{(\alpha)}(x,x')+G_F^{(D)}(s(x,x')).\label{eq:greendecomp}
\end{equation}

Equation \ref{eq:greendecomp} illustrates the break of  AdS invariance of the Green's function, as it may not depend on the geodetic interval $s$ entirely anymore.  We attribute the break on the maximal symmetry of $G_F$ to the imposition of different boundary conditions for each angular mode.

\section{Renormalized quantities for a conformal massless scalar field in $\mathrm{AdS}_4$}\label{sec:renormads}

In order to shed light on what we have discussed so far, we shall specialize to four spacetime dimensions, $\mathrm{AdS}_4$. For simplicity on the computation of quantities of interest, let us restrict ourselves to a conformally invariant, massless scalar field, $\phi$, i.e., $m=0$ and $\xi=\frac{1}{6}$.  In this case, from Eq.~\ref{eq:nu2}, we get $\nu=1/2$, and from Eq.~\ref{eq:sigma2}, we find that $\sigma=(2l+1)/2$.
Equation \ref{eq:waveop} becomes
\begin{equation}
    \left(-\frac{d^2}{d\rho^2}+\frac{l(l+1)}{\sin^2\rho}\right)f=\omega^2 f,
\end{equation}
and its solutions are
\begin{equation}
    f=\sqrt{\sin\rho}\left(\mathbf{C}_1\cdot P_{\omega-1/2}^{l+1/2}(\cos\rho)+\mathbf{C}_2\cdot Q_{\omega-1/2}^{l+1/2}(\cos\rho)\right),
\end{equation}
where $\mathbf{C}_1$ and $\mathbf{C}_2$ are
constants to be determined, and $P$ and $Q$ are the associated Legendre functions of the First and Second kinds, respectively.
Square integrability requires $f$ to fall off at the origin $\rho=0$, hence $\mathbf{C}_1\rightarrow0$\footnote{Formula $14.8.1$ of Ref.~\cite{abramo} shows the divergence of $P$ at $\rho=0$.}. A complete set of eigenfunctions is then
\begin{equation}
    f_{\omega,l}(\rho)=\mathcal{N}_{\omega,l}\cdot\sqrt{\sin\rho}\cdot Q_{\omega-1/2}^{l+1/2}(\cos\rho),
\end{equation}
for normalization constants $\mathcal{N}_{\omega,l}$ to be determined.

As discussed in Sec.~\ref{sec:scalarads}, boundary conditions at infinity are necessary to prescribe the dynamical evolution of the field in $\mathrm{AdS}_n$. In case $\nu=1/2$, Robin boundary conditions \ref{eq:Robinbc} are the appropriate ones. We aim to provide an example of the setups discussed in the last section. For that, we will consider that all non-spherically symmetric modes respect Dirichlet boundary conditions. However, the $l=0$ mode will be chosen to satisfy Robin condition with a parameter $\alpha$. As discussed above, the vacuum will not be $\mathrm{AdS}$ invariant in this case. However, since the non-trivial boundary condition is on $l=0$ mode, we still preserve spherical symmetry.

Formulas $14.5.3$ and $14.5.4$ in Ref.~\cite{abramo} allow us to describe the behavior of $f_{\omega, l}$ and its derivative at the boundary, as follows
\begin{eqnarray}
    f_{\omega,l}(\rho\rightarrow\pi/2)&\sim -\mathcal{N}_{\omega,l}\frac{2^{l-1/2}\sqrt{\pi}\sin\left(\frac{(l+\omega)\pi}{2}\right)\Gamma\left(\frac{l+\omega+1}{2}\right)}{\Gamma\left(\frac{-l+\omega+1}{2}\right)},\label{eq:fasympt} \\
    \frac{df_{\omega,l}}{d\rho}\bigg|_{\rho\rightarrow\pi/2}&\sim- \mathcal{N}_{\omega,l}\frac{2^{l+1/2}\sqrt{\pi}\cos\left(\frac{(l+\omega)\pi}{2}\right)\Gamma\left(\frac{l+\omega+2}{2}\right)}{\Gamma\left(\frac{-l+\omega}{2}\right)}\label{eq:dfasympt}
\end{eqnarray}
For $l>0$, all modes satisfy $f_{\omega,l}(\rho\rightarrow\pi/2)=0$ (Dirichlet boundary condition), thus its positive quantized frequencies are
\begin{equation}
    \omega=2r+l, \,r\in\mathbb{N}\cup\{0\}.
    \label{eq:quantdir}
\end{equation}
For $l=0$, we calculate the ratio between derivative \ref{eq:dfasympt} and function \ref{eq:fasympt} to use it in \ref{eq:Robinbc}, i.e.,
\begin{eqnarray}
    \left[\frac{df_{\omega,0}}{d\rho}\Big/f_{\omega,0}\right]_{\rho=\pi/2}&=2\cot\left(\omega\frac{\pi}{2}\right)\frac{\Gamma\left(1+\frac{\omega}{2}\right)}{\Gamma\left(\frac{\omega}{2}\right)}\nonumber\\
    &=\omega \cot\left(\omega\frac{\pi}{2}\right)=-\frac{1}{\alpha}.\label{eq:bcconform}
\end{eqnarray}
Positivity condition \ref{eq:positi} requires that
\begin{equation}
    \frac{1}{\alpha}\geq- \left|\frac{\Gamma(-1/2)}{\Gamma(1/2)}\right|\frac{\Gamma\left(1\right)^2}{\Gamma\left(1/2\right)^2}=-\frac{2}{\pi}\Rightarrow \alpha\leq-\frac{\pi}{2}\
    \mathrm{or} \ \alpha\geq 0.
\end{equation}
In our analysis, we consider $\alpha\geq0$, which includes Dirichlet, $\alpha=0$, and Neumann, $\alpha\rightarrow\infty$, cases.

\begin{figure}[b]
    \centering
    \includegraphics[width=.85\linewidth]{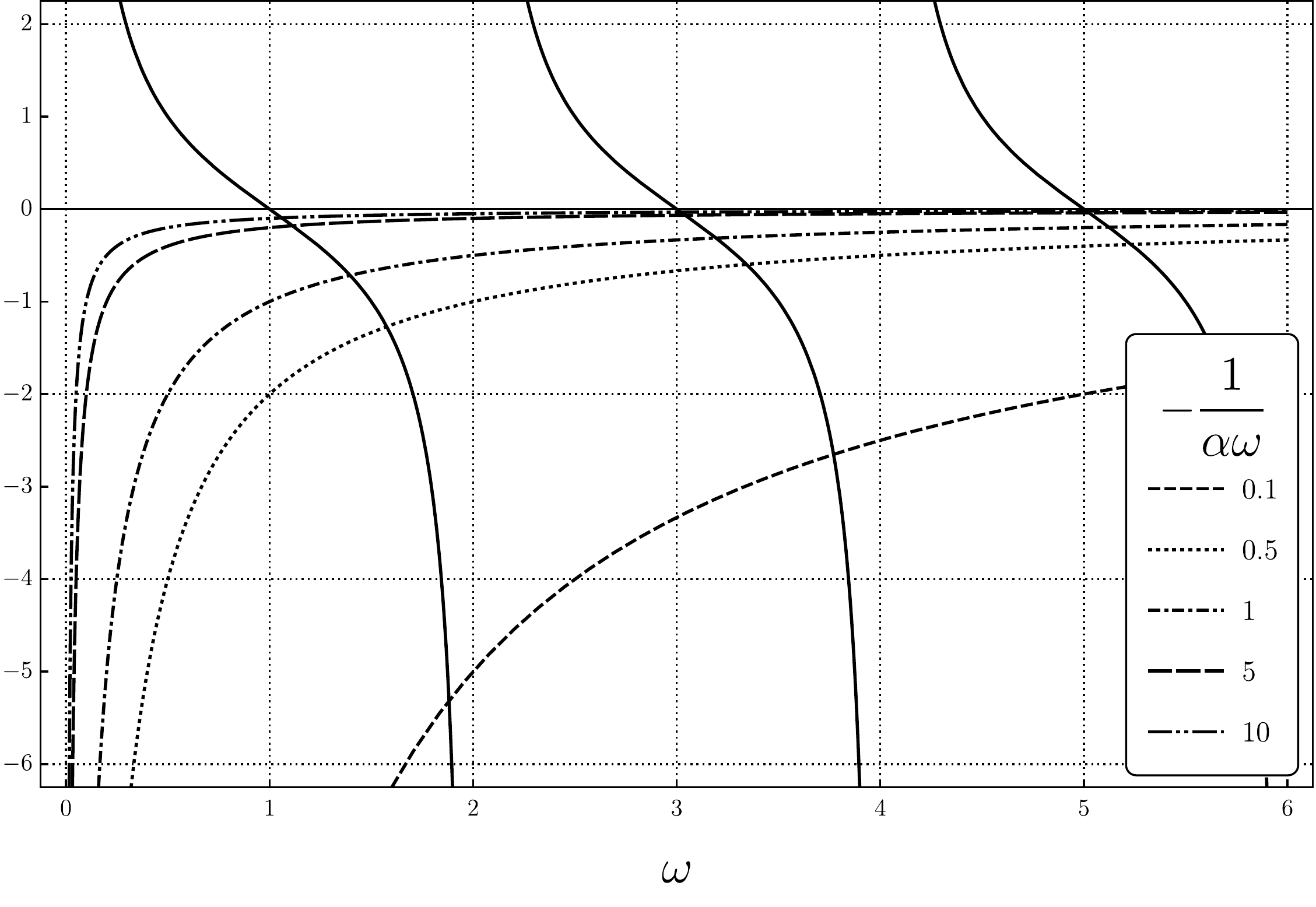}
    \caption{Quantization condition for $\omega$ imposed by \ref{eq:bcconform}. The solid lines show the function $\cot(\omega\pi/2)$ and all other curves are $-1/(\alpha\omega)$, for a few values of $\alpha$.}
    \label{fig:quantbc}
\end{figure}
Equation \ref{eq:bcconform} imposes a quantization condition for the frequencies $\omega$ in terms of the parameter $\alpha$. Except for $\alpha=0$ and $\alpha=\infty$, it cannot be solved analytically for an arbitrary value $\alpha$. One can readily verify that, in the Neumann case ($\alpha\rightarrow\infty$), the frequencies are odd integers. Meanwhile, for Dirichlet, they are even integers, which is consistent with Eq.~\ref{eq:quantdir}.

In our procedure, we employed the software Mathematica~\cite{mathematica} to solve equation \ref{eq:bcconform} numerically in a determined range of $\omega$ for several values of $\alpha$. As shown in Fig.~\ref{fig:quantbc}, the solutions of \ref{eq:bcconform} are given by the intersection points between the two functions. We can see that $\omega$ values for arbitrary $\alpha$ always lie between an odd number and its next even integer, which are precisely the frequencies for Neumann and Dirichlet conditions, respectively. Thus, given a Neumann frequency, $\omega_{N,r}=2r-1$, and a Dirichlet one, $\omega_{D,r}=2r$, for $r>0$, we may denote an $\alpha$ frequency between them as $\omega_{\alpha, r}$, even though it is not an integer number.

\subsection{Quadratic field fluctuations $\langle\phi^2\rangle$}\label{sec:phi2}
Before computing the Green's function, it is useful to write solution $f_{\omega_{\alpha, r},0}$ in a more convenient form and normalize it accordingly. Using Ref.~\cite{abramo}, we find\footnote{For convenience, we change the lower label in $f_{\omega,0}$ from $\omega_{\alpha, r}$ to $r$ simply, and add an upper index $\alpha$ to denote our choice of boundary condition.}
\begin{equation}
    f^{(\alpha)}_{r,0}(\rho)=H^{-1}\sqrt{\frac{2}{\omega_{\alpha, r}\pi-\sin(\omega_{\alpha, r}\pi)}}\sin(\omega_{\alpha, r}\rho).
\end{equation}
Now, we recall our discussion from last section to construct the appropriate Green's function. We can decompose our Green's functions in two parts, i.e.,
\begin{eqnarray}
  \fl  G_F^{(\alpha)}(x,x')=-i\frac{\cot\rho\cot\rho'}{4\pi H^2}\nonumber\\
    \qquad\times\sum_{r>0}\bigg( \frac{2}{\omega_{\alpha, r}\pi-\sin(\omega_{\alpha, r}\pi)}\sin(\omega_{\alpha, r}\rho)\sin(\omega_{\alpha, r}\rho')e^{-i\omega_{\alpha, r}(\tau-\tau')}\nonumber\\
   \qquad\qquad\qquad-\frac{2}{2r\pi}\sin(2r\rho)\sin(2r\rho')e^{-i2r(\tau-\tau')}\bigg),\label{eq:greenalpha1}
\end{eqnarray}
and 
\begin{eqnarray}
  \fl  G_F^{(D)}(x,x')=-iH^{-2}\cot\rho\cot\rho'\nonumber\\
  \times\sum_{r\geq0}\sum_{l\geq0}  \sum_{m=-l}^l |\mathcal{N}_{r, l}|^2Y^m_l(\theta_j,\varphi)[Y^m_l(\theta_j',\varphi')]^*f_{r, l}(\rho)f_{r, l}(\rho')e^{-i2r(\tau-\tau')}.\label{eq:greendir}
\end{eqnarray}

Our `Dirichlet' Green's function \ref{eq:greendir} is obtained from a summation of AdS invariant modes of the wave equation. Hence, it respects maximal symmetry and recovers the results of Burgess and L\"{u}tken, $G_F^{(BL)}$, and Allen and Jacobson, $G_F^{(AJ)}$, i.e., $G_F^{(D)}(x,x')\equiv G_F^{(D)}(s(x,x'))$. As Kent and Winstanley show in \cite{kent}, approaching the coincidence limit $s\rightarrow0$, the function $G_F^{(D)}$ diverges according to the Hadamard form. Thus, point-splitting renormalization can be employed to compute finite quantities. Furthermore, they obtain the Hadamard forms in $\mathrm{AdS}$ for any spacetime dimension through a systematic method, based on \cite{deca}.

In the particular case of $\mathrm{AdS}_4$, for a conformally invariant field, the Green's function $G_F^{(D)}$ has the Hadamard form given by
\begin{equation}
    G_F^{(D)}(s)\sim -\frac{i}{4\pi^2s^2}, \quad s\rightarrow0.
\end{equation}
After renormalization, it may be written as~\cite{kent}
\begin{equation}
    \Big[G_F^{(D)}\Big]_\mathrm{ren}(s)=-\frac{i}{8\pi^2H^2}\left\{-\frac{1}{6}+\frac{13}{240}\frac{s^2}{H^2}+\mathcal{O}(s^4)\right\}.
\end{equation}
We may find the expectation value of the quadratic field fluctuations as follows
\begin{equation}
    \langle\phi^2\rangle^{(D)}=i\lim_{s\rightarrow0}\Big[G_F^{(D)}\Big]_\mathrm{ren}(s)=-\frac{1}{48\pi^2H^2},\label{eq:phi2dir}
\end{equation}
which is naturally in accordance with the results in Ref.~\cite{kent}. Analogously, the effect of our Green's function \ref{eq:greenalpha1} on $\langle\phi^2\rangle$ appears when taking the coincidence limit $x'\rightarrow x$. However, calculating $G_F^{(\alpha)}$ analytically is impossible, since the summation is taken over numerical values of frequencies. Hence, we adopt a numerical approach to find our results.

We expect $G_F^{(\alpha)}$ to be finite, since the Hadamard form took care of the divergences in $G_F^{(D)}$. On the other hand, we cannot perform the infinite sum in \ref{eq:greenalpha1} numerically, so a residual divergent behavior might appear. Through our computations, we noted it was convenient to take the coincidence limit in the radial coordinate first, i.e., $\rho'\rightarrow\rho$, and then in the time coordinate. Thus, our final step would be to take the limit of $\tau'\rightarrow\tau$. It is more convenient though, to analytically extend the function on the complex plane and take the limit through the imaginary axis, i.e., $\tau'\rightarrow\tau+i\epsilon$, hence $\tau-\tau'\rightarrow-i\epsilon$. Finally, by multiplying $G_F^{(\alpha)}$ by $i$, we will have an entirely real-valued function that, in the limit $\epsilon\rightarrow0$, yields directly the quadratic fluctuations of the field, and it is much simpler for us to handle it numerically.

Before implementing the numerical routine, we considered the only case that can be treated analytically, which is the Neumann condition, $\alpha\rightarrow\infty$. In this situation, the frequencies are $\omega_{\infty,r}=2r-1$, for $r>0$, and the Green's function reduces to the following summation
\begin{equation}
     \fl   iG_F^{(\infty)}(\epsilon,\rho,\rho)=\frac{\cot^2\rho}{2\pi^2H^2}
        \times\sum_{r>0}\bigg( \frac{\sin^2((2r-1)\rho)}{2r-1}e^{-(2r-1)\epsilon}    -\frac{\sin^2(2r\rho)}{2r}e^{-2r\epsilon}\bigg),\label{eq:greenalpha2}
\end{equation}
which we calculated using Mathematica~\cite{mathematica}, resulting
\begin{equation}
 \fl  iG_F^{(\infty)}(\epsilon,\rho,\rho)=\frac{\cot^2\rho}{16\pi^2H^2}\times\log\left[\cosh\left(\frac{\epsilon}{2}\right)^4\sec\left(\rho-i\frac{\epsilon}{2}\right)^2\sec\left(\rho+i\frac{\epsilon}{2}\right)^2\right].
\end{equation}
It is straightforward to find the expectation value $\langle\phi^2\rangle^{(N)}$ by simply taking $\epsilon\rightarrow0$, i.e.,
\begin{equation}
    \langle\phi^2\rangle^{(N)}(\rho)=\frac{\cot^2\rho}{4\pi^2H^2}\log\left[\sec\left(\rho\right)\right].\label{eq:phi2neumann}
\end{equation}
The function $\langle\phi^2\rangle^{(N)}$ is finite because both terms inside the sum in Eq.~\ref{eq:greenalpha2} diverge with same strength. Naturally, their subtraction eliminates the infinities. In particular, the last term in Eq.~\ref{eq:greenalpha2}, the Dirichlet counterpart of $G_F^{(\alpha)}$, denoted $G_F^{(\alpha ,D)}$, appears for all values of $\alpha$ and dictates the divergent behavior at $\epsilon\rightarrow0$. We find its form by calculating the infinite summation and expanding it in powers of $\epsilon$, i.e.,
\begin{equation}
    iG_F^{(\alpha ,D)}(\epsilon,\rho)=\frac{\cot^2\rho}{8\pi^2H^2}\left\{-\log\epsilon+\log[\sin(2\rho)]+\mathcal{O}(\epsilon^2)\right\}.\label{eq:greendiverg}
\end{equation}

Our numerical approach to find the expectation value $\langle\phi^2\rangle^{(\alpha)}$ proceeded as follows:
\begin{enumerate}
    \item Given a value for $\alpha$, solve Eq.~\ref{eq:bcconform} to find the frequencies $\omega_{\alpha,r}$ up to $r_{\mathrm{max}}=5000$;
    \item Given a value of $\rho$ between $0$ and $\pi/2$, compute numerically the truncated summation
    \begin{equation}
          \fl  iG_F^{(\alpha)}(\epsilon,\rho,\rho)\approx\frac{\cot^2\rho}{2\pi}\sum_{r=1}^{r_{\mathrm{max}}} \bigg( \frac{\sin^2(\omega_{\alpha, r}\rho)}{\omega_{\alpha, r}\pi-\sin(\omega_{\alpha, r}\pi)}e^{-\omega_{\alpha, r}\epsilon}
    -\frac{\sin^2(2r\rho)}{2r\pi}e^{-2r\epsilon}\bigg)
    =:\texttt{f}^{(\alpha)}_\rho[\epsilon],
    \end{equation}
for $50$ values of $\epsilon$ equally spaced in the range $0.002$ to $0.1$.\footnote{Our choice for $r_{\mathrm{max}}$ and the range of $\epsilon$ was made so the last term of the sum would be negligible with respect to the first one. Indeed, the first term is of order $e^{-2\cdot1\cdot 0.002}\sim 10^{-1}$, while the last is $e^{-2\cdot5000\cdot0.002}\sim 10^{-9}$. Also, we needed $\epsilon$ small enough so the divergent behavior would appear.}
    \item Fit the function $\texttt{f}^{(\alpha)}_\rho[\epsilon]$ using a model that reproduces the divergent behavior in Eq.~\ref{eq:greendiverg} followed by a Taylor expansion up to order $\epsilon^2$, i.e.,
    \begin{equation}
        \texttt{f}[\epsilon]=\texttt{a}+\texttt{b}\log[\epsilon]+\texttt{c}\cdot\epsilon+\texttt{d}\cdot\epsilon^2.
    \end{equation}
    As $G_F^{(\alpha)}$ is a finite quantity, we expect the divergent behavior of $\texttt{f}^{(\alpha)}_\rho[\epsilon]$ to be extremely attenuated. We have found coefficients $\texttt{b}$ ranging between $10^{-9}$ and $10^{-12}$, recovering the expected \emph{almost-finite} behavior. The coefficients $\texttt{c}$ and $\texttt{d}$ were effective on reducing the residuals of the fit. Finally, $\texttt{a}$ gives the approximated finite numerical value of $\langle\phi^2\rangle^{(\alpha)}$ at the point $\rho$.
    \item Repeat steps $2$ and $3$ for as many values of $\rho$ between $0$ and $\pi/2$ as desired.
    \item Repeat the entire procedure for another value of $\alpha$.
\end{enumerate}

\begin{figure}[h!]
    \centering
    \includegraphics[width=.9\linewidth]{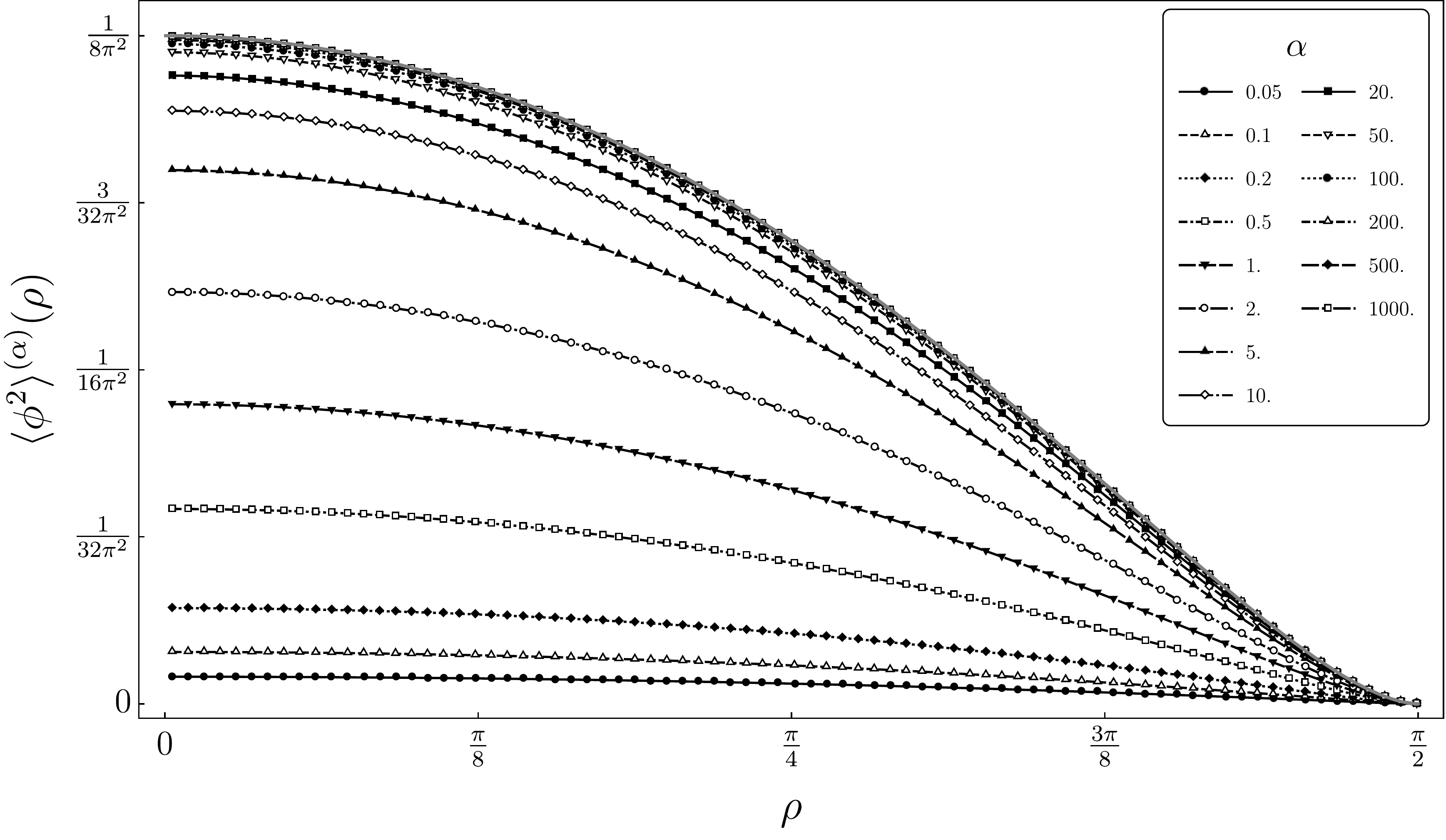}
    \caption{Contribution to the expectation value of the quadratic field fluctuations due to Robin boundary conditions at infinity for the spherically symmetric mode. The solid gray curve shows the analytical solution of the Neumann case, $\langle\phi^2\rangle^{(N)}$. $H$ is set to one.}
    \label{fig:phi2alpha}
\end{figure}

We followed the scheme described above for $14$ values for the parameter $\alpha$. We chose $80$ equally spaced points in the range $(0,\pi/2)$ to obtain a good resolution of the behavior of $\langle\phi^2\rangle^{(\alpha)}(\rho)$. Our results are plotted in Fig.~\ref{fig:phi2alpha}. The curve corresponding to $\alpha=1000$ reproduces almost perfectly the analytic Neumann result \ref{eq:phi2neumann}. Accordingly, as we approach the other extreme, $\alpha=0$ - corresponding to Dirichlet conditions - we can see the curves getting closer to zero. Consistently, if $\alpha=0$, then $G_F^{(\alpha)}$ indeed vanishes, as one can see from Eq.~\ref{eq:greenalpha1}. 

\subsection{Energy-momentum tensor fluctuations $\langle T^\mu_{\ \nu}\rangle^{(\alpha)}$}\label{sec:emt}

In~\cite{kent}, the authors obtain the renormalized energy-momentum tensor $\langle T^\mu_{\ \nu}\rangle_\mathrm{ren}$ in $\mathrm{AdS}_n$. They use the formula from Ref.~\cite{deca} 
\begin{equation}
    \fl \braket{T_{\mu\nu}}_\mathrm{ren}=-[G]_{\mu\nu}+\frac{1}{2}(1-2\xi)[G]_{;\mu\nu}
    +\frac{1}{2}\left(2\xi-\frac{1}{2}g_{\mu\nu}\nabla_\sigma\nabla^\sigma[G]+\xi R_{\mu\nu}[G]\right)+\Theta_{\mu\nu},\label{eq:emtren}
\end{equation}
where
\begin{eqnarray}
    \qquad[G](x)&:=\lim_{x'\rightarrow x}i\left[G_F\right]_\mathrm{ren}(x,x'),\label{eq:functiong}
    \\  \qquad  [G]_{\mu\nu}(x)&:=\lim_{x'\rightarrow x}i\left[G_F\right]_\mathrm{ren}(x,x')_{;\mu\nu},\label{eq:functiongmn}
\end{eqnarray}
and $\Theta_{\mu\nu}$ is a purely geometric tensor constructed to be conserved. Kent and Winstanley find that the non-geometrical component of the tensor is proportional to the metric tensor, which is completely consistent with the maximal symmetry of $\mathrm{AdS}$. In our particular case of a conformally invariant field in four spacetime dimensions, we have
\begin{equation}
    \langle T_{\mu\nu}\rangle^{(D)}_\mathrm{ren}=-\frac{1}{960\pi^2H^4}g_{\mu\nu},\label{eq:emtdir}
\end{equation}
and the geometric tensor $\Theta_{\mu\nu}$ is identically zero.
We may obtain this renormalized expectation value from Green's function $\Big[G_F^{(D)}\Big]_\mathrm{ren}$, hence is associated with Dirichlet conditions in all modes of the wave equation.

Here, we want the contributions to the energy-momentum tensor coming from $G^{(\alpha)}_F$. Our approach will be analogous to that of the Green's functions: we decompose $\langle T^\mu_{\ \nu}\rangle_\mathrm{ren}$ into two parts, one carrying the boundary condition, denoted $\langle T^\mu_{\ \nu}\rangle^{(\alpha)}_\mathrm{ren}$ with $16$ components $\mathrm{T}^\mu_{\ \nu}$, and another one reproducing the Dirichlet results as in Eq.~\ref{eq:emtdir}.

In our case, equations \ref{eq:functiong} and \ref{eq:functiongmn} may be written as
\begin{equation}
    [G](\rho)=\lim_{\epsilon\rightarrow0}iG_F^{(\alpha)}(\epsilon,\rho,\rho),
\end{equation}
and
\begin{eqnarray}
    [G]_{\mu\nu}(\rho)&=\lim_{\epsilon\rightarrow0}i(G_F^{(\alpha)})_{;\mu\nu}(\epsilon,\rho,\rho)\nonumber\\
    &=\lim_{\epsilon\rightarrow0}i\left[(G_F^{(\alpha)}),_{\mu\nu}-\Gamma^\lambda_{\mu\nu}(G_F^{(\alpha)}),_{\lambda}\right],
\end{eqnarray}
from which it follows that $[G](\rho)\equiv\braket{\phi^2}^{(\alpha)}(\rho)$. According to formula \ref{eq:emtren}, we have here
\begin{equation}
    \langle T_{\mu\nu}\rangle^{(\alpha)}_\mathrm{ren}=-[G]_{\mu\nu}+\frac{1}{3}[G]_{;\mu\nu}
    -\frac{1}{12}\nabla_\kappa\nabla^\kappa [G]g_{\mu\nu}-\frac{1}{2H^2}[G]g_{\mu\nu}.\label{eq:emtalpha}
\end{equation}
Considering all non-vanishing Christoffel symbols, the definitions for $[G]$ and $[G]_{\mu\nu}$, and the symmetric condition $\langle T_{\mu\nu}\rangle^{(\alpha)}_\mathrm{ren}=\braket{T_{\nu\mu}}^{(\alpha)}_\mathrm{ren}$, we readily verify that the only non-vanishing components are diagonal terms and the term $\mathrm{T}_{\tau\rho}\ (=\mathrm{T}_{\rho\tau})$. Let us recall the temporal inversion ($\tau\rightarrow-\tau$) symmetry of $\mathrm{AdS}$, denoted $\mathrm{I}$, given in four dimensions by the transformation matrix $\mathrm{I}_\mu^{\mu'}=\mathrm{diag}(-1,1,1,1)$. As none of our quantities depend explicitly on $\tau$, we expect this discrete symmetry to be preserved. In particular, we expect $\mathrm{T_{\tau x^j}}=\mathrm{T_{-\tau x^j}}=\mathrm{T_{\tau' x'^j}}$, for $x^j=(\rho,\theta,\varphi)$. On the other hand, $\langle T^\mu_{\ \nu}\rangle^{(\alpha)}_\mathrm{ren}$ transforms as a tensor, so we have
\begin{equation}
    \langle T_{\mu'\nu'}\rangle^{(\alpha)}_\mathrm{ren}=\mathrm{I}^\mu_{\mu'}\mathrm{I}^\nu_{\nu'}\langle T_{\mu\nu}\rangle^{(\alpha)}_\mathrm{ren}
    \Rightarrow\mathrm{T_{-\tau \rho}}=\mathrm{T_{\tau' \rho'}}=\mathrm{I}^\tau_{\tau'}\mathrm{I}^\rho_{\rho'}\mathrm{T_{\tau \rho}}=-\mathrm{T_{\tau \rho}}.
\end{equation}
That yields $\mathrm{T_{\tau \rho}}=-\mathrm{T_{\tau \rho}}$, which then implies $\mathrm{T_{\tau \rho}}=\mathrm{T_{\rho\tau}}\equiv0$.

At this point, we have a diagonal tensor, whose remaining components may be calculated using Eq.~\ref{eq:emtalpha}. Our computational efforts were not successful when trying to compute the numerical expressions directly. However, we came up with a solution based on some properties that $\langle T^\mu_{\ \nu}\rangle^{(\alpha)}_\mathrm{ren}$ must satisfy, based on the definition of $\langle T^\mu_{\ \nu}\rangle_\mathrm{ren}$.

Let us first consider the effect of the trace anomaly. One can readily verify that it is respected by $\langle T^\mu_{\ \nu}\rangle^{(D)}_\mathrm{ren}$~\cite{kent,deca}, i.e.,
\begin{equation}
    \braket{T^{\mu}_{\ \mu}}_\mathrm{ren}=\braket{T^{\mu}_{\ \mu}}^{(D)}_\mathrm{ren}=-\frac{1}{240\pi^2H^4},
\end{equation}
so our tensor $\langle T^\mu_{\ \nu}\rangle^{(\alpha)}_\mathrm{ren}$ must be traceless,
\begin{equation}
    \braket{T^{\mu}_{\ \mu}}^{(\alpha)}_\mathrm{ren}\equiv 0=\mathrm{T}_{\ \tau}^{ \tau}+\mathrm{T}_{\ \rho}^{ \rho}+\mathrm{T}_{\ \theta}^{ \theta}+\mathrm{T}_{\ \varphi}^{ \varphi},\label{eq:traceless}
\end{equation}
which is our first constrain on the remaining diagonal components. We may use the symmetries of $\mathrm{AdS}$ as well. Although our Green's function breaks  $\mathrm{AdS}$ invariance of the radial coordinate $\rho$, all other symmetries should remain valid. In $\mathrm{AdS}_4$ there exist $10$ Killing fields corresponding to the following isometries: one temporal translation, three rotations, four boosts and four spatial translations. From which, we only expect the first two to be preserved after imposing Robin boundary conditions in only one of the modes.

The temporal Killing field, $t=\partial_\tau$, yields a conservation equation along with its flow, given by the Lie derivative of the tensor with respect to $t$, i.e.,
\begin{equation}
    \mathcal{L}_t\braket{T^{\mu}_{ \ \nu}}^{(\alpha)}_\mathrm{ren}=0\Rightarrow t^\sigma\partial_\sigma \braket{T^{\mu}_{ \ \nu}}^{(\alpha)}_\mathrm{ren}=\partial_\tau\braket{T^{\mu}_{ \ \nu}}^{(\alpha)}_\mathrm{ren}=0,
\end{equation}
which shows that all components of $\braket{T^{\mu}_{ \ \nu}}^{(\alpha)}_\mathrm{ren}$ are independent of $\tau$. Additionally, we have the generators of spherical symmetry, given by the following Killing fields
\begin{eqnarray}
    \chi_1&=\partial_\varphi,\\
    \chi_2&=\cos\varphi\partial_\theta-\cot\theta\sin\varphi\partial_\varphi,\\
    \chi_3&=-\sin\varphi\partial_\theta-\cot\theta\cos\varphi\partial_\varphi.
\end{eqnarray}
Since a combination of them is still a Killing field, we may use $\chi_2$ and $\chi_3$ to obtain $\chi_4=\partial_\theta$. We can use $\chi_1$ and $\chi_4$ to find other two conservation equations similar to that of $t$, as follows
\begin{eqnarray}
    \mathcal{L}_{\chi_1}\braket{T^{\mu}_{ \ \nu}}^{(\alpha)}_\mathrm{ren}=0\Rightarrow \chi_1^\sigma\partial_\sigma \braket{T^{\mu}_{ \ \nu}}^{(\alpha)}_\mathrm{ren}=\partial_\varphi\braket{T^{\mu}_{ \ \nu}}^{(\alpha)}_\mathrm{ren}=0,\\
    \mathcal{L}_{\chi_4}\braket{T^{\mu}_{ \ \nu}}^{(\alpha)}_\mathrm{ren}=0\Rightarrow \chi_4^\sigma\partial_\sigma \braket{T^{\mu}_{ \ \nu}}^{(\alpha)}_\mathrm{ren}=\partial_\theta\braket{T^{\mu}_{ \ \nu}}^{(\alpha)}_\mathrm{ren}=0.
\end{eqnarray}
These equations show us that $\braket{T^{\mu}_{ \ \nu}}^{(\alpha)}_\mathrm{ren}$ can be a function of $\rho$ only, i.e., $\braket{T^{\mu}_{ \ \nu}}^{(\alpha)}_\mathrm{ren}\equiv\braket{T^{\mu}_{ \ \nu}}^{(\alpha)}_\mathrm{ren}(\rho)$.

Finally, the conservation equation,
\begin{equation}
    \nabla^\nu \braket{T^\mu_{\ \nu}}_\mathrm{ren}=0,\label{eq:conserv}
\end{equation}
provide us with the last set of constrains. As $\langle T^\mu_{\ \nu}\rangle^{(D)}_\mathrm{ren}$ is proportional to the metric, it is automatically conserved, since $\nabla^\mu g_{\mu\nu}=0$. Hence, for $\langle T^\mu_{\ \nu}\rangle_\mathrm{ren}$ to be entirely conserved, we must impose Eq.~\ref{eq:conserv} on $\langle T^\mu_{\ \nu}\rangle^{(\alpha)}_\mathrm{ren}$ as well, which, using the properties we have found for $\braket{T^{\mu}_{ \ \nu}}^{(\alpha)}_\mathrm{ren}$ so far, reduces to
\begin{eqnarray}
    \partial_\rho\mathrm{T}^{\rho}_{\ \rho}-\tan\rho\mathrm{T}^{\tau}_{\ \tau}+(4\csc(2\rho)+\tan\rho)\mathrm{T}^{\rho}_{\ \rho}\nonumber\\
    \hspace{90pt}-2\csc(2\rho)(\mathrm{T}^{\theta}_{\ \theta}+\mathrm{T}^{\varphi}_{\ \varphi})=0,\label{eq:conserv1}\\
    \cot\theta(\mathrm{T}^{\theta}_{\ \theta}-\mathrm{T}^{\varphi}_{\ \varphi})=0\Rightarrow \mathrm{T}^{\theta}_{\ \theta}=\mathrm{T}^{\varphi}_{\ \varphi}.\label{eq:conserv2}
\end{eqnarray}

Before discussing our numerical approach for the expectation value of the energy-momentum tensor, we treat the case $\alpha\rightarrow\infty$, i.e., Neumann boundary condition. Again, we were able to find an analytic result only in this situation. We used equation \ref{eq:emtalpha} to find the formulas for components,\footnote{We are setting $H=1$ to clear the expressions, later on we reinsert it.}

\begin{eqnarray}
\mathrm{T}^{\tau}_{\ \tau}=\cos^2\rho\left ( [G]_{\tau\tau}-\tan\rho[G]_{r} \right )+\frac{1}{3}\cos^2\rho\tan\rho [G],_\rho\nonumber \\ \hspace{100pt}-\frac{1}{12}\left ( \cos^2\rho [G],_{\rho\rho}+2\cot\rho[G],_\rho \right )-\frac{1}{2}[G],
\\
\mathrm{T}^{\rho}_{\ \rho}=-\cos^2\rho\left ( [G]_{\rho\rho}-\tan\rho[G]_{r} \right )+\frac{1}{3}\cos^2\rho\left ( [G],_{\rho\rho}-\tan\rho[G],_{r} \right )\nonumber\\
 \hspace{140pt} -\frac{1}{12}\left ( \cos^2\rho [G],_{\rho\rho}+2\cot\rho[G],_\rho \right )-\frac{1}{2}[G],
\\
\mathrm{T}^{\theta}_{\ \theta}=\mathrm{T}^{\varphi}_{\ \varphi}=-\cot^2\rho\tan\rho[G]_{r} +\frac{1}{3}\cot^2\rho\tan\rho [G],_\rho\nonumber\\
 \hspace{100pt}-\frac{1}{12}\left ( \cos^2\rho [G],_{\rho\rho}+2\cot\rho[G],_\rho \right )-\frac{1}{2}[G].
\end{eqnarray}
Our attempts to compute $[G]_{\rho\rho}$ and $[G]_{\rho}$ analytically and numerically were not successful. Hence, we adopted another approach that combined the explicit formulas above and the constrains given by equations \ref{eq:traceless} and \ref{eq:conserv1}.

Let us conveniently define a function $\mathrm{F}(\rho)$ depending exclusively on the quantities we were able to compute, namely $[G]_{\tau\tau}(\rho)$ and $[G](\rho)$, as follows
\begin{eqnarray}
    \mathrm{F}(\rho)&:=\csc^2\rho \mathrm{T}^{\tau}_{\ \tau}(\rho)-\mathrm{T}^{\theta}_{\ \theta}(\rho)\nonumber\\& \ =\cot^2\rho
\left\{[G]_{\tau\tau}-\frac{1}{12}\left ( \cos^2\rho [G],_{\rho\rho}+2\cot\rho[G],_\rho \right )-\frac{1}{2}[G]\right\}.\label{eq:defF}
\end{eqnarray}
Using Eq.~\ref{eq:traceless} and recalling that $\mathrm{T}^{\theta}_{\ \theta}=\mathrm{T}^{\varphi}_{\ \varphi}$, we find that
\begin{equation}
    \mathrm{T}^{\rho}_{\ \rho}(\rho)=2\mathrm{F}(\rho)-(1+2\csc^2\rho) \mathrm{T}^{\tau}_{\ \tau}(\rho),\label{eq:Trr}
\end{equation}
and applying it to \ref{eq:conserv1}, we have
\begin{equation}
   \partial_\rho\mathrm{T}^{\tau}_{\ \tau}+2\frac{9-\cos(2\rho)+2\csc^2\rho}{(5-\cos(2\rho))\cot\rho}\mathrm{T}^{\tau}_{\ \tau}
    =2\frac{(\sin(2\rho)\mathrm{F}'+(7-\cos(2\rho))\mathrm{F})}{(5-\cos(2\rho))\cot\rho}.\label{eq:diffTtt}
\end{equation}
The equation above is of the form
\begin{equation}
    u'(\rho)+p(\rho)u(\rho)=q(\rho),
\end{equation}
upon the identifications $u\equiv\mathrm{T}^{\tau}_{\ \tau}$,
\begin{equation}
    p(\rho)=2\frac{9-\cos(2\rho)+2\csc^2\rho}{(5-\cos(2\rho))\cot\rho}
\end{equation}
and
\begin{equation}
    q(\rho)=2\frac{(\sin(2\rho)\mathrm{F}'(\rho)+(7-\cos(2\rho))\mathrm{F}(\rho))}{(5-\cos(2\rho))\cot\rho}.
\end{equation}
 One can verify that
 \begin{equation}
     u(\rho)=\exp\left[-\int d\rho \ p\right]\left(\int_0^\rho d\rho'\  \exp\left[\int d\rho' p\right]\cdot q + \mathbf{C} \right)
 \end{equation}
 solves the equation. In our case, we have
 \begin{equation}
     \exp\left[\int d\rho \ p\right]=\tan\rho\sec^3\rho\sqrt{5-\cos(2\rho)},
 \end{equation}
 which vanishes at $\rho=0$ and diverges at $\rho=\pi/2$. Naturally, the inverse function $\exp\left[-\int d\rho \ p\right]$ vanishes at the boundary, but diverges at $\rho=0$ with strength $1/\rho$. As it is physically reasonable to ask for a finite $\mathrm{T}^{\tau}_{\ \tau}$ at $\rho=0$, we set $\mathbf{C}$ to zero. Finally, we compute $\mathrm{T}^{\tau}_{\ \tau}$ using the following expression
  \begin{equation}
    \fl \mathrm{T}^{\tau}_{\ \tau}(\rho)=2\frac{\cot\rho\cos^3\rho}{\sqrt{5-\cos(2\rho)}}\int_0^\rho d\rho'\  \frac{\tan^2\rho'\Big(\sin(2\rho')\mathrm{F}'(\rho')+(7-\cos(2\rho'))\mathrm{F}(\rho')\Big)}{\cos^3\rho'\sqrt{5-\cos(2\rho')}}.\label{eq:Ttt} 
 \end{equation}

For the Neumann case, we used our previous analytic results and found $\mathrm{F}$ to be
\begin{equation}
    \mathrm{F}(\rho)=\frac{\cot^2\rho}{48\pi^2}\Big(\csc^2\rho+2+2(\csc^4\rho-1)\log(\sec\rho)\Big).
\end{equation}
Applying it in Eq.~\ref{eq:Ttt}, and then using \ref{eq:Trr} and \ref{eq:defF}, we find
\begin{eqnarray}
  \fl  \braket{T^{\mu}_{ \ \nu}}^{(N)}_\mathrm{ren}(\rho)=\frac{\cot^2\rho}{48\pi^2H^4}\bigg\{\Big(\sin^2\rho\Big)\mathrm{diag}(1,-1,0,0)\nonumber\\
    \hspace{100pt}+\Big(1-2\cot^2\rho\log(\sec\rho)\Big)\mathrm{diag}(1,1,-1,-1)\bigg\}.
\end{eqnarray}
Now, we have a result to compare our numerical ones with.

To compute the function $\mathrm{F}$ numerically, we used our previous results of $\braket{\phi^2}^{(\alpha)}$($=[G]$), but we also need $[G]_{\tau\tau}$. According to \ref{eq:functiongmn}, we find it by taking the second derivative of $G_F^{(\alpha)}(\tau,\tau',\rho,\rho)$ with respect to $\tau$ and, then, taking the coincidence limit. In the convention we adopted, $\partial_{\tau\tau}=-\partial_{\epsilon\epsilon}$. Its then expected that the divergent behavior of the Dirichlet counterpart $G_F^{(\alpha,D)}$ is not that of Eq.~\ref{eq:greendir} anymore. Indeed, we find it to be
\begin{equation}
    -\partial_{\epsilon\epsilon}G_F^{(\alpha,D)}=\frac{\cot^2\rho}{8\pi^2}\\
    \times\left\{ -\frac{1}{\epsilon^2}-\frac{1}{24}(5+\cos(4\rho))\csc^2\rho\sec^2\rho+\mathcal{O}(\epsilon^2)\right\}.
\end{equation}

Our numerical procedure to find the expectation value of the energy-momentum tensor fluctuations was:
\begin{enumerate}
    \item Given a value for $\alpha$, use the frequencies $\omega_{\alpha,r}$ found before;
    \item Given a value of $\rho$ between $0$ and $\pi/2$, compute numerically the truncated summation
    \begin{equation}
        \fl-i\partial_{\epsilon\epsilon}G_F^{(\alpha)}\approx-\frac{\cot^2\rho}{2\pi}\sum_{r=1}^{r_{\mathrm{max}}} \bigg( \frac{\omega_{\alpha, r}^2\sin^2(\omega_{\alpha, r}\rho)}{\omega_{\alpha, r}\pi-\sin(\omega_{\alpha, r}\pi)}e^{-\omega_{\alpha, r}\epsilon}
    -\frac{(2r)^2\sin^2(2r\rho)}{2r\pi}e^{-2r\epsilon}\bigg),
    \end{equation}
    denoted $\texttt{F}^{(\alpha)}_\rho[\epsilon]$, for $50$ values of $\epsilon$ equally spaced in the range $0.002$ to $0.1$.
    \item Fit the function $\texttt{F}^{(\alpha)}_\rho[\epsilon]$ using a model that reproduces the divergent behavior followed by a Taylor expansion up to order $\epsilon^2$, i.e.,
    \begin{equation}
        \texttt{h}[\epsilon]=\texttt{a}+\frac{\texttt{b}}{\epsilon^2}+\texttt{c}\cdot\epsilon+\texttt{d}\cdot\epsilon^2.
    \end{equation}
    As expected, the divergent behavior of $\texttt{F}^{(\alpha)}_\rho[\epsilon]$ is extremely attenuated, and the coefficient $\texttt{b}$ is negligible compared to the others. Again, the coefficients $\texttt{c}$ and $\texttt{d}$ were effective on reducing the residuals of the fit. Finally, $\texttt{a}$ gives the finite approximated numerical value of $[G]_{\tau\tau}$ at the point $\rho$.
    \item Repeat steps $2$ and $3$ for as many values of $\rho$ between $0$ and $\pi/2$ as desired to obtain the complete $[G]_{\tau\tau}(\rho)$.
    \item Use our previous results for $[G]$ together with $[G]_{\tau\tau}$ in Eq.~\ref{eq:defF} to find a numerical interpolation of $\mathrm{F}(\rho)$, denoted $\texttt{F}[\rho]$.
    \item Given a value of $\rho$ between $0$ and $\pi/2$, use $\texttt{F}[\rho]$ in Eq.~\ref{eq:Ttt} and perform a numerical integration to obtain an approximate value of $\mathrm{T}_{\ \tau}^{\tau}$ at that specific $\rho$.
    \item Repeat step 6 for several values of $\rho$ to find a complete numerical function $\mathrm{T}_{\ \tau}^{\tau}$. With that in hands, compute $\mathrm{T}_{\ \rho}^{\rho}$ and $\mathrm{T}_{\ \theta}^{\theta}$ using equations \ref{eq:Trr} and \ref{eq:defF}.
    \item Repeat the entire procedure for a different value of $\alpha$.
\end{enumerate}
\begin{figure}[h!]
    \centering
    \includegraphics[width=.9\linewidth]{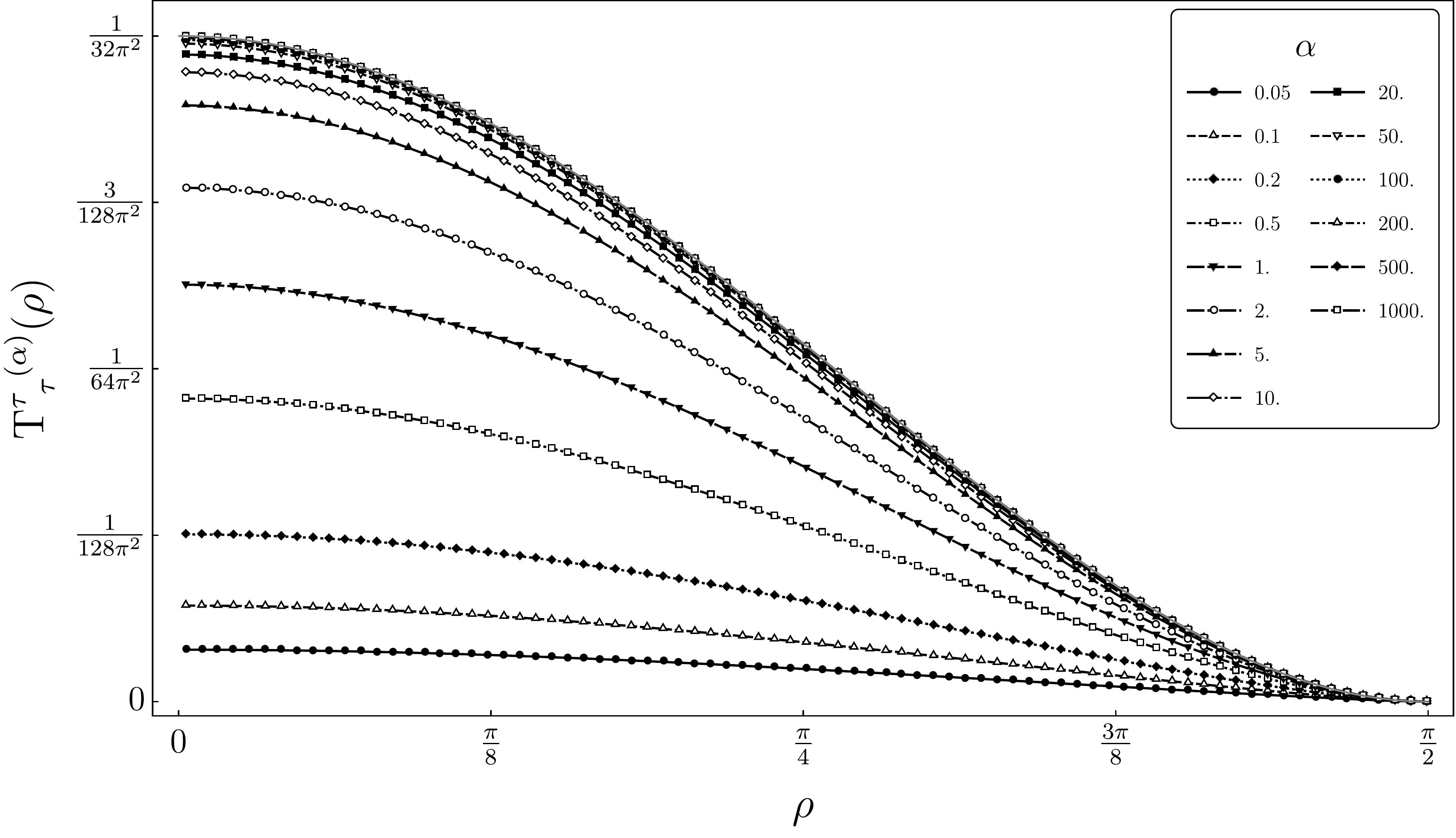}
    \caption{Contribution to the expectation value $\mathrm{T}_{\ \tau}^{\tau}$ due to Robin boundary conditions at infinity for the spherically symmetric mode. $H$ is set to one.}
    \label{fig:Tttalpha}
\end{figure}

Similarly to our results for the expectation value of the field squared, we followed the numerical procedure for $14$ values of $\alpha$. We have found all components of $\langle T^\mu_{\ \nu}\rangle^{(\alpha)}_\mathrm{ren}$. In Fig.~\ref{fig:Tttalpha}, we can see $\mathrm{T}_{\ \tau}^{\tau}$ for several values of $\alpha$, it is clear that the form of the function follows the analytic result for the Neumann condition (plotted in gray).

\section{\label{sec:disc}Discussion and further remarks}

Avis, Isham, and Storey took a first-step, in Ref.~\cite{avis}, towards the development of a quantum field theory in anti-de Sitter spacetime. They acknowledged that the conformal infinity poses a serious causality issue to the wave equation but solve it by regulating the information flow through the boundary `by hand.' They imposed the so-called `transparent' and `reflective' boundary conditions at infinity in analogy to a box in Minkowski spacetime. In this way, they quantized the fields in the Einstein Static Universe and restricted it to the AdS later. 

Conversely, in this article, we considered the developments made by Ishibashi and Wald in \cite{ishibashi1}, where they propose a physically consistent prescription for the dynamical evolution of fields. In the particular case that we have considered, they show that the imposition of mixed boundary conditions at the spatial infinity is sufficient to determine the evolution of quantum fields uniquely. 

In the setup studied by Kent and Winstanley, in Ref.~\cite{kent}, all angular modes of the wave equation satisfy the same Dirichlet boundary condition at infinity. Their results are consistent with the maximal symmetry of AdS. Hence, the expectation values of field-dependent quantities fluctuate in the same way throughout spacetime, i.e., they are coordinate-independent. In the light of Wald's and Ishibashi's developments, we presented a setup here that puts up to question how necessary it is to impose the same boundary conditions to all modes of the wave equation. Indeed, we are not aware of any requirement of nature that precludes us from considering various setups in terms of boundary conditions.

\begin{figure}[h]
    \centering
    \includegraphics[width=.9\linewidth]{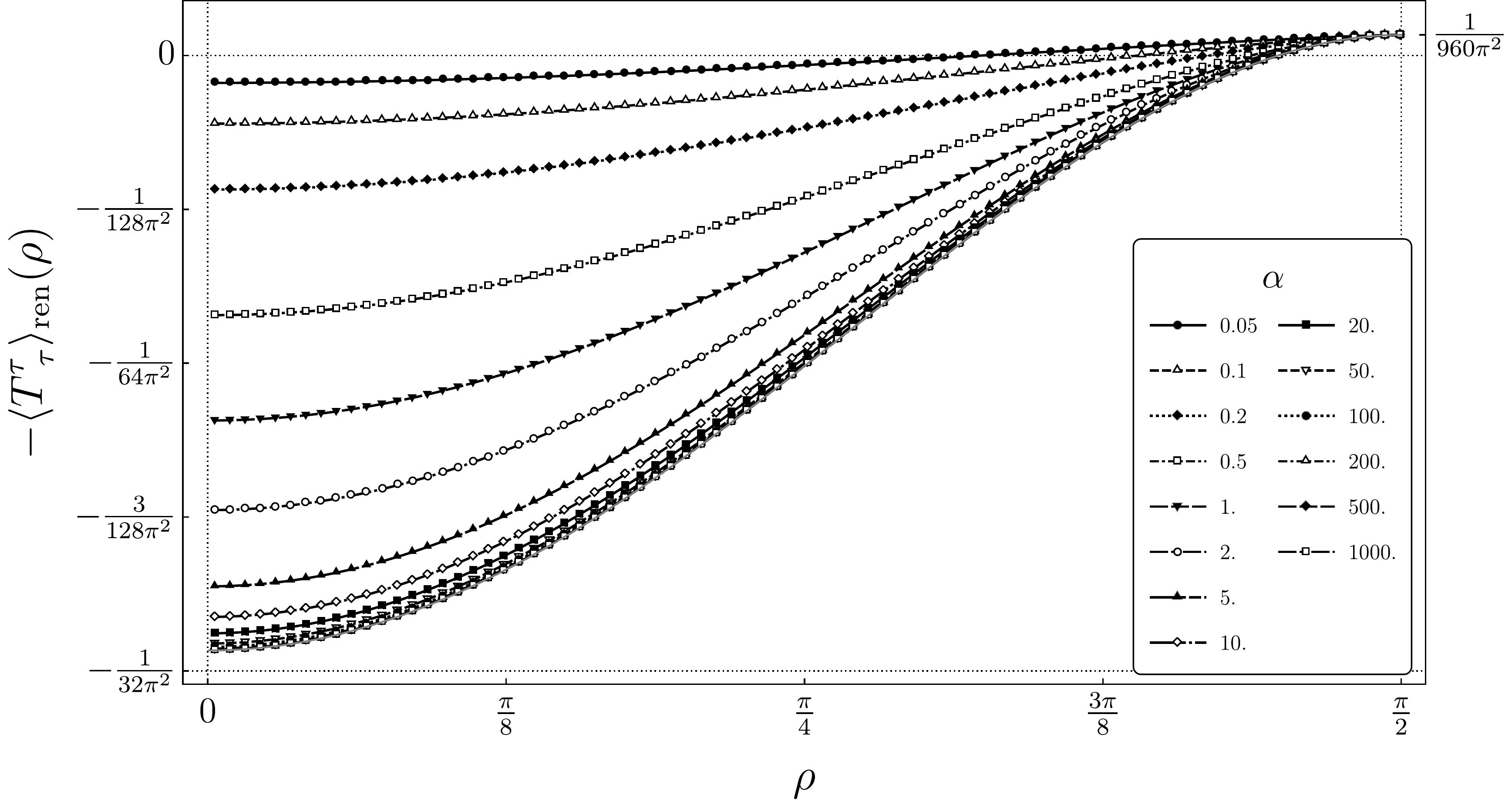}
    \caption{Energy densitiy $-\langle T_{\ \tau}^{\tau}\rangle_\mathrm{ren}$ of a massless scalar field conformally coupled to $\mathrm{AdS}_4$ for several Robin boundary conditions. $H$ is set to one.}
    \label{fig:minusTtt}
\end{figure}

Our analysis indicated a violation of AdS invariance in the Green's functions, which carried out implications on the related quantities: the quadratic fluctuations of the field and the energy-momentum tensor. Both of them are now dependent on the radial coordinate for any values of the parameter $\alpha$, as shown in Fig.~\ref{fig:phi2alpha} and \ref{fig:Tttalpha}.  At this stage, any attempt of obtaining a back-reacted metric using Einstein's semi-classical equations,
\begin{equation}
G_{\mu\nu}=8\pi G \langle T_{\mu\nu} \rangle_\mathrm{ren}\equiv 8\pi G\left(\braket{T_{\mu\nu}}^{(D)}_\mathrm{ren}+\braket{T_{\mu\nu}}^{(\alpha)}_\mathrm{ren}\right),
\end{equation}
would not yield a maximally symmetric metric anymore, but a spherically symmetric one. In these conditions, the coordinate system used to define the angular modes of the wave equation will be privileged. In particular, in this system, the energy density reaches its minimum at the origin $\rho=0$, as shown in Fig.~\ref{fig:minusTtt}. 

In Fig.~\ref{fig:minusTtt}, we can see a clear violation of the weak energy condition in most of the spacetime, except close to the boundary, where the Dirichlet contribution, $-\langle T_{\ \tau}^{\tau}\rangle^{(D)}_\mathrm{ren}$, pushes the energy density back up over zero. Even though such violation is no stranger to us - as can be observed in the Casimir effect -  it appeared as a consequence of the contribution from the Robin boundary condition exclusively. Indeed, the Dirichlet term, $-\langle T_{\ \tau}^{\tau}\rangle^{(D)}_\mathrm{ren}$, of the energy density is positive throughout the entire spacetime. Thus, it is safe to assert that the violation of the weak energy condition is a direct consequence of the imposition of non-Dirichlet boundary conditions at infinity.

\ack

V.~S. Barroso and J.~P.~M. Pitelli are grateful to Professor R.~M. Wald and G.~Satishchandran for enlightening discussions and also thank the Enrico Fermi Institute for the kind hospitality. J.~P.~M. Pitelli thanks Funda\c c\~ao de Amparo \`a Pesquisa do Estado de S\~ao Paulo (FAPESP) (Grant No. 2018/01558-9). V.~S. Barroso thanks FAPESP (Grant No. 2018/09575-0). Finally, we all thank FAPESP (Grant No. 2013/09357-9).


\section*{References}
\begin{thebibliography}{99}

\bibitem{maldacena}
J. Maldacena, {\it The Large-N Limit of Superconformal Field Theories and Supergravity}, IInt. J. Theor. Phys. {\bf 38}, 1113 (1999).

\bibitem{kent}
C. Kent and E. Winstanley, {\it Hadamard Renormalized Scalar Field Theory on Anti–de Sitter Spacetime}, Phys. Rev. D {\bf 91}, 044044 (2015).

\bibitem{avis}
S. J. Avis, C. J. Isham and D. Storey, {\it Quantum Field Theory in Anti-de Sitter Space-time}, Phys. Rev. D {\bf 18}, 11 (1978).

\bibitem{wald}
R.~M. Wald, {\it Dynamics in Nonglobally Hyperbolic, static space-times}, J. Math. Phys. {\bf 21}, 2802 (1980).

\bibitem{ishibashi1}
A.~Ishibashi and R.~M. Wald, {\it 
Dynamics in Non-Globally-Hyperbolic Static Spacetimes II: General analysis of prescriptions for dynamics}, Class. Quant. Grav. {\bf 20}, 3815 (2003).


\bibitem{ishibashi2}
A.~Ishibashi and R.~M. Wald, {\it 
Dynamics in Non-Globally-Hyperbolic Static Spacetimes III: Anti-de Sitter spacetime}, Class. Quant. Grav. {\bf 21}, 2981 (2004).

\bibitem{wald1}
 R.~M. Wald, {\it General Relativity}, (University of Chicago Press, Chicago, 1984).
 
 
\bibitem{birrel}
N.~D. Birrel and P.~C.~W. Davis, {\it Quantum Fields in Curved Space}, (Cambridge University Press, 1982).


\bibitem{hawking}
S. W. Hawking and G. F. R. Ellis, {\it The Large Scale Structure of Spacetime}, (Cambridge University Press, 1973). 

\bibitem{reed1}
M. Reed and B. Simon, {\it Methods of Modern Mathematical Physics I: Functional Analysis}, (Elsevier Science, 1975).

\bibitem{reed2}
M. Reed and B. Simon, {\it Methods of Modern Mathematical Physics II: Fourier Analysis, Self-Adjointness}, (Elsevier Science, 1981).


\bibitem{allen}
B. Allen and T. Jacobson, {\it Vector Two-Point Functions in Maximally Symmetric Spaces}, Comm. Math. Phys. {\bf 103}, 669 (1986).

\bibitem{burgess}
C. P. Burgess and C. A. L\"{u}tken, {\it Propagators and Efective Potentials in Anti-de Sitter Space}, Phys. Lett. B {\bf 153}, 137 (1985).


\bibitem{abramo}
M. Abramowitz and I.~A. Stegun, {\it Handbook of Mathematical Functions}, (Washington, DC, 1972).


\bibitem{mathematica}
Wolfram Research, Inc., Mathematica, Version 12.0, Champaign, IL (2019).


\bibitem{deca}
Y. Décanini and A. Folacci, {\it Hadamard Renormalization of the Stress-Energy Tensor for a Quantized Scalar Field in a General Spacetime of Arbitrary Dimension}, Phys. Rev. D {\bf 78}, 044025 (2008).

\end{thebibliography}
\end{document}